\documentclass[aps,prb,twocolumn,showpacs,preprintnumbers,amsmath,amssymb,superscriptaddress]{revtex4}%
\usepackage{amssymb}

\usepackage{graphicx}
\usepackage{dcolumn}
\usepackage{bm}
\usepackage{color}
\begin{document}


\title{Thermal and Quantum critical Properties of overdoped La$_{2-x}$Sr$_{x}$CuO$_{4}$}

\author{T. Schneider}
\email{tschnei@physik.uzh.ch}
\affiliation{Physik-Institut der Universit\"{a}t Z\"{u}rich, Winterthurerstrasse 190, CH-8057, Switzerland}

\begin{abstract}
We analyze the extended superfluid density data of Bo\v{z}ovi\'{c} \textit{%
et al}. taken on homogenous thin films to explore the critical properties of
the thermal (TSM) and quantum superconductor to metal (QSM) transitions in
overdoped La$_{2-x}$Sr$_{x}$CuO$_{4}$. The temperature dependence reveals
remarkable agreement with d-wave BCS behavior, where $\sigma \propto
1/\lambda ^{2}$. No sign of the expected KTB transition is observable. We
show that the critical amplitude $\rho _{s0}=\rho _{s}\left( T\right)
/\left( 1-T/T_{c}\right) $ scales as $T_{c}\propto \rho _{s0}^{1/2}$ and
with that as $T_{c}\propto \rho _{s}\left( 0\right) $, empirically verified
by Bo\v{z}ovi\'{c} \textit{et al}. Together with additional evidence for BCS
behavior in overdoped La$_{2-x}$Sr$_{x}$CuO$_{4}$ we show that the scaling
relation $T_{c}\propto \lambda \left( 0\right) ^{-1}\propto \Delta \left(
0\right) \propto \xi \left( 0\right) ^{-1}\propto H_{c2}\left( 0\right)
^{-1/2}\propto \lambda _{0}^{-1}\propto \Delta _{0}\propto \xi
_{0}^{-1}\propto H_{c20}^{-1/2}$ applies, by approaching the QSM and TSM
transitions. This differs from the dirty limit behavior $\lambda \left(
0\right) \propto \xi \left( 0\right) ^{1/2}$, maintains the $T_{c}$ \
independence of the Ginzburg-Landau parameter $\kappa =\lambda _{o}/\xi
_{0}$, and puts a stringent constraint on the effect at work. We notice
that potential candidates appear to be the nonlocal effects on the
penetration depth of clean d-wave superconductors explored by Kosztin and
Legget.

\end{abstract}

%
\maketitle

\begin{figure}[htb]
\includegraphics[scale=0.4]{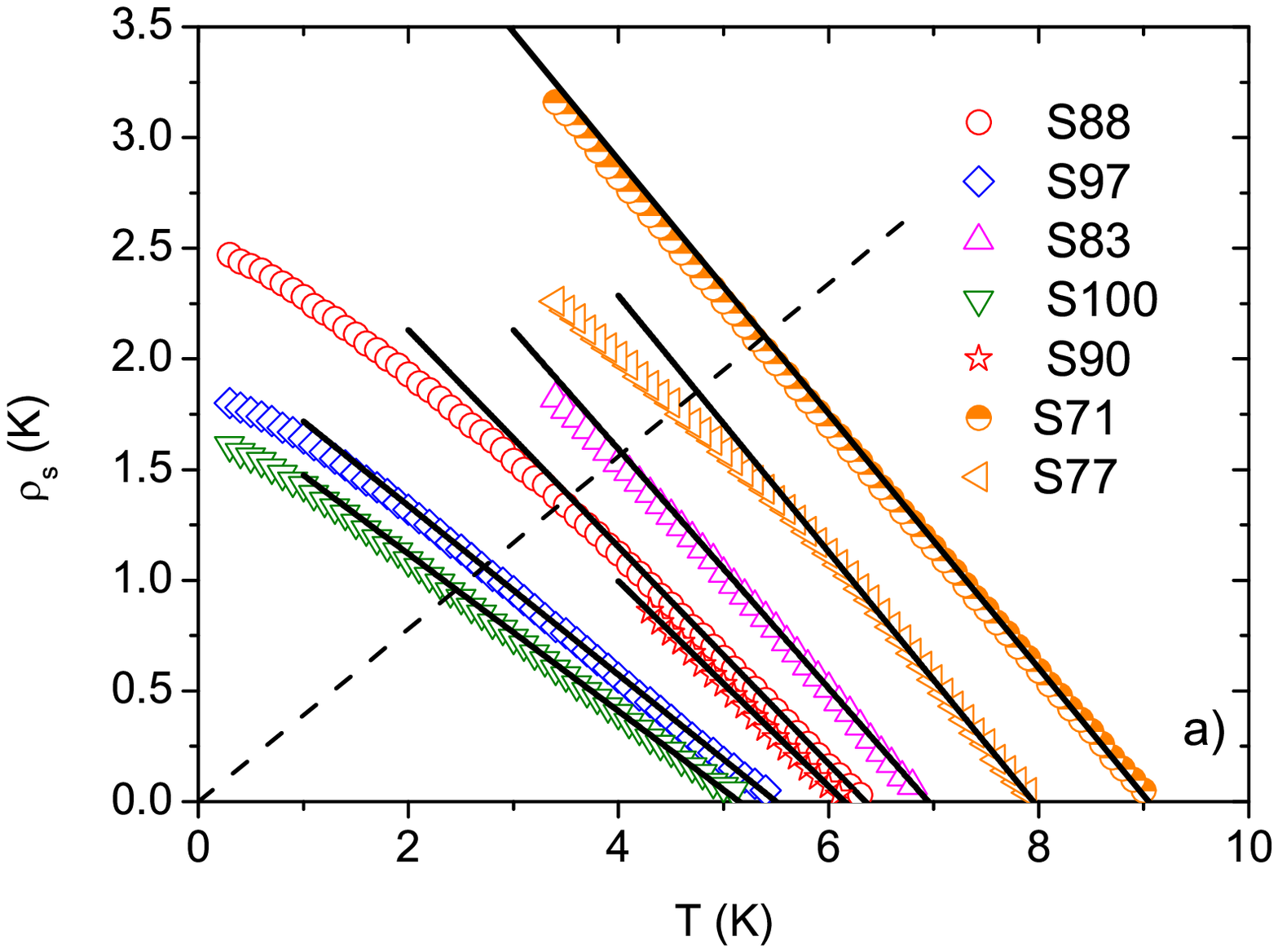}
\includegraphics[scale=0.4]{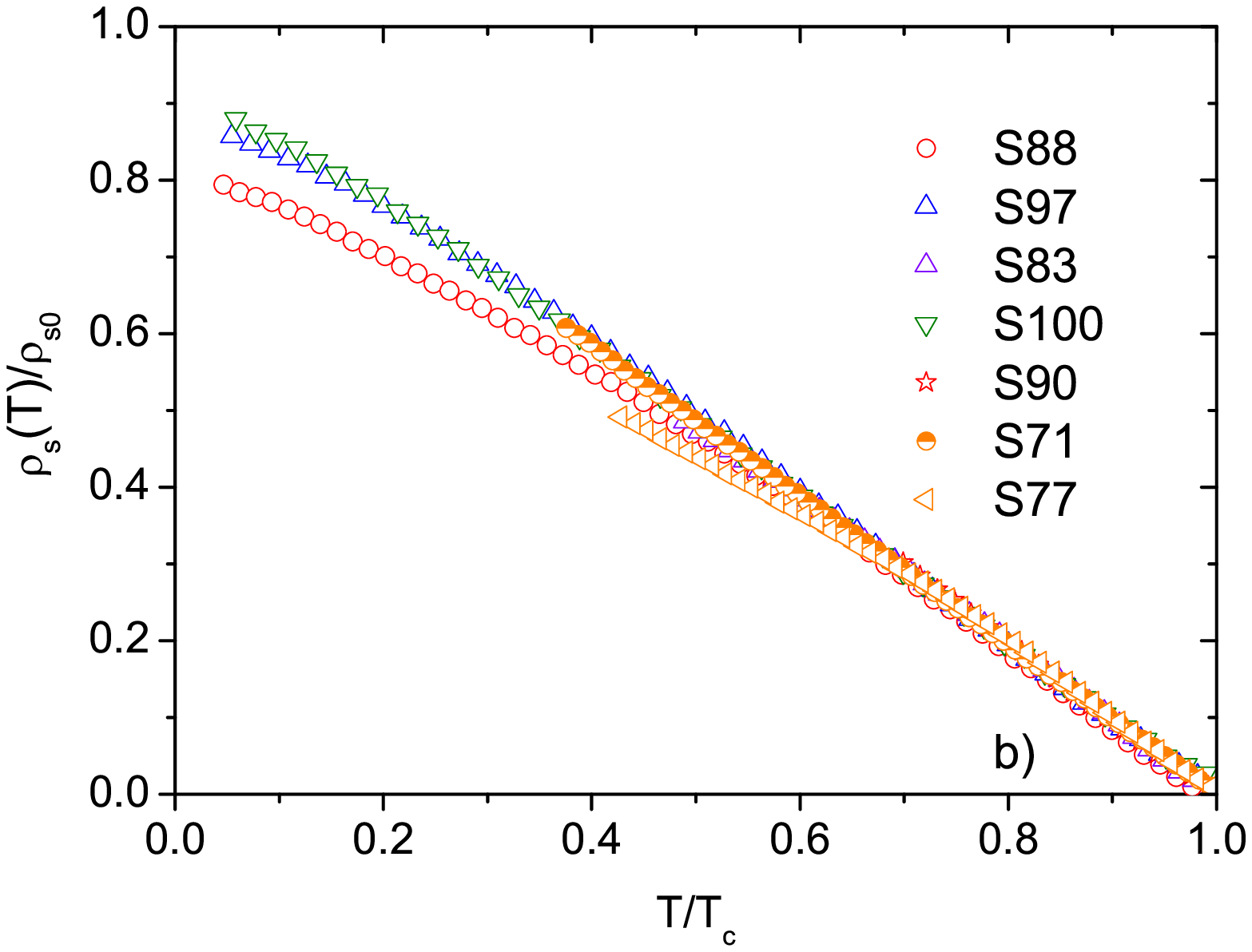}
\caption{a) Superfluid stiffness $\rho _{s}\left( T\right) $ for a selection
of overdoped thin films taken from Bo\v{z}ovi\'{c} \textit{et.al}. \cite{bosovic}. 
S denotes the selected samples. The dashed line marks the KTB
transition temperature where the universal jump in the superfluid stiffness
should occur (Eq. (\ref{eq3})). The solid lines are fits to Eq. (\ref{eq4})
yielding for $\rho _{s0}$ and $T_{c}$ the estimates collected in Fig. (\ref%
{fig2}). b) Scaling plot $\rho _{s}\left( T\right) /\rho _{s0}$ vs. $T/T_{c}.$}
\label{fig1}
\end{figure}

\begin{figure}[htb]
\includegraphics[scale=0.4]{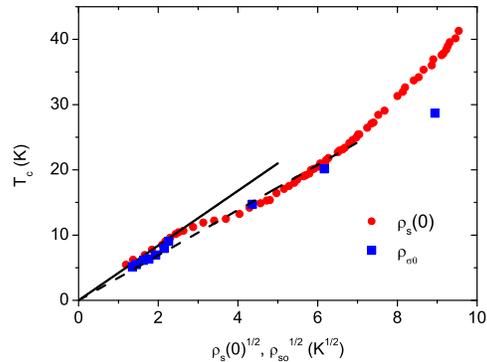}
\caption{Estimates for the critical amplitude $\rho _{s0}\left( T_{c}\right) $%
, shown as $T_{c}$ versus $\rho _{s0}^{1/2}\propto 1/\lambda _{0}^{1/2}$,
derived from the fits shown in Fig. \ref{fig1}a. The dashed line is $%
T_{c}=f\rho _{s0}^{1/2}$ with $f=3.46$ K$^{1/2}$. For comparison we included
$T_{c}$ versus $\rho _{s}\left( 0\right) ^{1/2}$ of \ Bo\v{z}ovi\'{c}
\textit{et al}. \cite{bosovic} $\rho _{s}\left( 0\right) \propto 1/\lambda
\left( 0\right) ^{2}$ is the zero temperature superfluid stiffness and the
solid line is $T_{c}=g\rho _{s}\left( 0\right) ^{1/2}$ with $g=4.2$ K$^{1/2}$.}
\label{fig2}
\end{figure}

Considering thin films and adopting the phase transition point of view the
thermal (TSP) and quantum (QSM) superconductor to metal transitions are
expected to fall onto the respective two dimensional (2D) xy- universality
class. Accordingly the TSP transition should exhibit the characteristic
Kosterlitz-Thouless-Berezinski (KTB) behavior \cite{kosterlitz},
including a discontinuous drop in the superfluid stiffness $\rho _{s}(T)\propto
d/\lambda ^{2}\left( T\right) $ from \cite{nelson},
\begin{equation}
\frac{d}{\lambda ^{2}\left( T_{c}^{-}\right) }=\frac{32\pi ^{2}}{\Phi
_{0}^{2}}k_{B}T_{c}\simeq 1.017T_{c},  \label{eq1}
\end{equation}%
to zero, with $d/\lambda ^{2}\left( T_{c}^{-}\right) $ in cm$^{-1}$ and $%
T_{c}$ in K. $\Phi _{0}=hc/2e\simeq 2.07\times 10^{-7}$erg$^{1/2}$cm$^{1/2}$%
. $d$ \ denotes the film thickness. In a homogenous film this relationship
applies if $d/\lambda ^{2}\left( T\right) $ is measured at zero frequency
and $\lambda ^{2}\left( T\right) /d$ is large compared to the lateral extent
of the film. Otherwise there is a rounded BKT-transition. In particular, if
the lateral extent of the homogeneous regions $L$ is finite the correlation
length cannot grow beyond $L$. Similarly, there is no phase transition at
finite frequency because the frequency scales as $1/\omega \propto \xi
^{z_{cl}}\propto L^{z_{cl}}$ and so gives rise to a smeared
Nelson-Kosterlitz jump \cite{book}. Noting that $d/\lambda ^{2}\left(
T\right) $ and $\rho _{s}(T)$ are related by%
\begin{equation}
\frac{d}{\lambda ^{2}\left( T\right) }=\frac{4\pi \alpha k_{B}}{\hbar c}\rho
_{s}(T),  \label{eq2}
\end{equation}%
where k$_{B}$ is Boltzmann`s , $\hbar $ Planck`s, $\alpha $ the fine
structure constant, and $c$ speed of light. From Eqs.(\ref{eq1}) and (\ref%
{eq2}) we obtain for the BKT line the relation%
\begin{equation}
T_{c}=\frac{\Phi _{0}^{2}\alpha }{8\pi \hbar c}\rho _{s}(T)\simeq 0.39\rho
_{s}(T),  \label{eq3}
\end{equation}%
with $T_{c}$ and $\rho _{s}$ in K. Fig. \ref{fig1}a shows data taken from Bo%
\v{z}ovi\'{c} \textit{et al}. \cite{bosovic} of the superfluid stiffness $%
\rho _{s}\left( T\right) $ for a selection of overdoped thin La$_{2-x}$Sr$%
_{x}$CuO$_{4}$. S denotes the selected samples. The dashed line is the KTB
transition temperature where at the intersection with $\rho _{s}\left(
T\right) $ the universal jump should occur. Obviously there are no signs for
the universal jump at the respective KTB transition temperatures. In fact,
the rather sharp TSM transitions occur considerably above the respective KTB
transition temperatures and reveal remarkable consistency with the
mean-field temperature dependence%
\begin{eqnarray}
\rho _{s}(T) &=&\rho _{s0}\left( T_{c}\right) t=\frac{4\pi \alpha k_{B}}{%
\hbar c}\frac{d}{\lambda ^{2}\left( T\right) }  \notag \\
&=&\frac{4\pi \alpha k_{B}}{\hbar c}\frac{d}{\lambda _{0}^{2}\left(
T_{c}\right) }t=\frac{4\pi \alpha k_{B}}{\hbar c}\frac{d}{\lambda
_{0}^{2}\left( T_{c}\right) }t,  \label{eq4}
\end{eqnarray}%
where $t=(1-T/T_{c})$. Note that such an extended linear $t$ dependence is
a characteristic of feature of d-wave BCS superconductors where
thermal phase fluctuations, driving the BKT transition, are fully neglected
\cite{prozorov}. The solid lines in Fig. \ref{fig1}a are fits to Eq. (\ref%
{eq4}). The resulting fit parameters, $\rho _{s0}$ and $T_{c}$, are
collected in Fig. (\ref{fig2}) and yield the scaling plot shown in Fig. \ref%
{fig1}b. Obviously, when the QSM transition is reached, the data fall onto a
straight line and thus confirm the BCS d-wave scenario very
impressively.
Given the evidence for the irrelevance of phase fluctuations, another
unexpected empirical fact emerges from Fig. \ref{fig2}, depicting $%
T_{c}\left( \rho _{s0}^{1/2}\right) $. The dashed line $T_{c}=f\rho
_{s0}^{1/2}$, describing the approach to the QSM transition rather well,
agrees with the zero temperature counterpart $T_{c}\propto \rho \left(
0\right) ^{1/2}$, verified by Bo\v{z}ovi\'{c} \textit{et al}. \cite{bosovic}.
In fact, this relationship contradicts the general belief that the
measured penetration depth corresponds to the London penetration depth,
whereupon%
\begin{equation}
\rho _{s}\left( 0\right) \propto 1/\lambda \left( 0\right) ^{2}\propto n,
\label{eq5}
\end{equation}%
applies, where $n$ is fully determined by the shape of the Fermi surface
\cite{einzel}. For circular or spherical Fermi surfaces it corresponds to
the electron density in the normal state. As shown by angle-resolved
photoemission experiments, this is not the case in overdoped La$_{2-x}$Sr$%
_{x}$CuO$_{4}$ \cite{razzoli}. However, the result is determined by the
shape of the Fermi surface and independent of $T_{c}$. Therefore even the
more general treatment of the zero temperature penetration depth is
incompatible with the empirical relation
\begin{eqnarray}
T_{c} &\propto &\rho _{s}(0)^{1/2}\propto \rho _{s0}^{1/2}  \notag \\
&\propto &1/\lambda \left( 0\right) \propto 1/\lambda _{0},  \label{eq6}
\end{eqnarray}%
emerging from Fig. \ref{fig2}. $T_{c}\propto \rho _{s}(0)^{1/2}$ was
verified by Bo\v{z}ovi\'{c} \textit{et al}. \cite{bosovic} and $T_{c}\propto
\rho _{s0}^{1/2}$ follows from our analysis shown in Fig. \ref{fig1},
yielding $\rho _{s0}\left( T_{c}\right) $ depicted in Fig.\ref{fig2}.
Note that the observation of a diminishing $\rho _{s}(0)$ in the overdoped regime
is confirmed by earlier measurements. \cite{niedermayer,bernhard,locquet,pana,lemberger}
 In addition there is the hyperscaling prediction \cite{book, kim}%
\begin{equation}
T_{c}\propto \rho _{s}(0)^{z/(D+z-2)},  \label{eq7}
\end{equation}%
where $z$ denotes the dynamic critical exponent of the quantum transition.
It is applicable whenever a phase transition line $T_{c}\left( x\right) $
exhibits a critical endpoint at $x=x_{c}$ of the tuning parameter $x$. Here
the transition temperature vanishes and a quantum phase transition occurs.
In thin films ($D=2$) it yields $T_{c}\propto \rho _{s}\left( 0\right) $,
irrespective of the value of the dynamic critical exponent $z$, and with
that it contradicts the empirical relation (\ref{eq5}).
\begin{figure}[htb]
\includegraphics[scale=0.4]{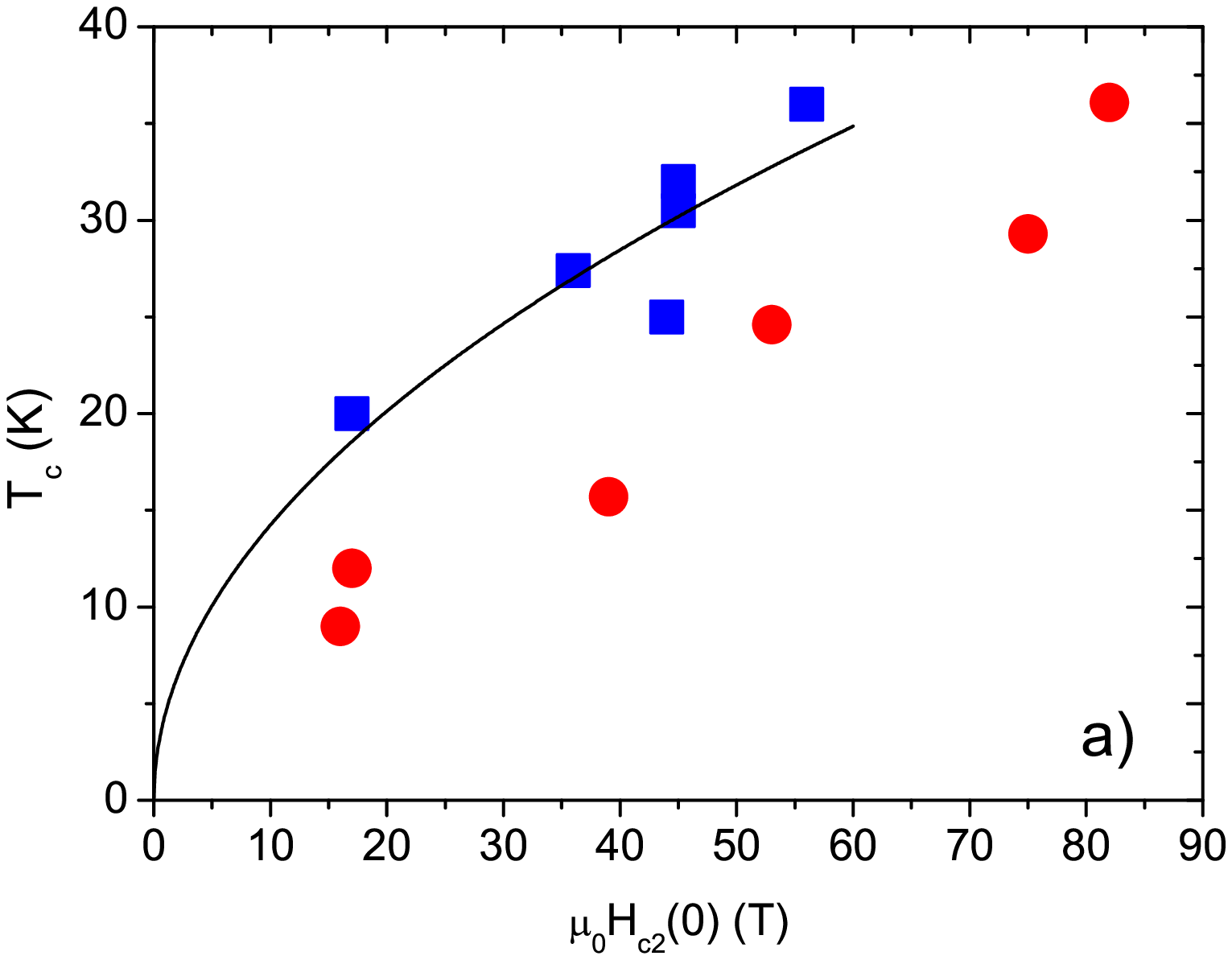}
\includegraphics[scale=0.4]{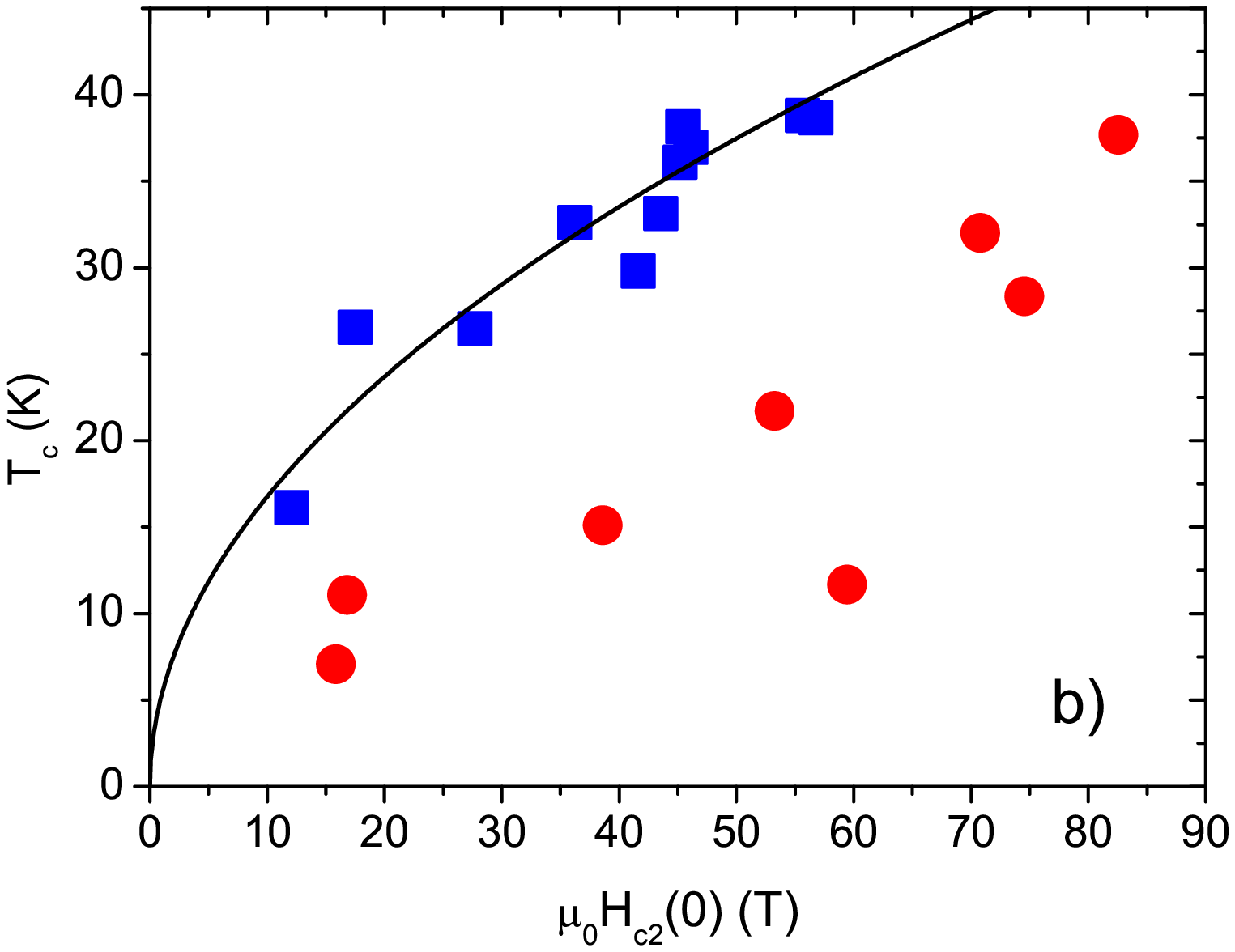}
\caption{ $T_{c}$ versus $H_{c2}(0)$. a) Taken from Y. Wang and H.-H. Wen as
derived from specific heat measurements on La$_{2-x}$Sr$_{x}$CuO$_{4}$ single
crystals \cite{wangwen}. ($\blacksquare $: overdoped regime; $\bullet $ :
underdoped regime). The line is $T_{c}=aH_{c2}(0)^{1/2}$ with $a=4.5KT^{-1/2}
$. b) Taken from Rourke \textit{et al.} as derived from magnetoresistance
measurements La$_{2-x}$Sr$_{x}$CuO$_{4}$ single crystals \cite{rourke}. ($%
\blacksquare $:overdoped regime; $\bullet $: underdoped regime). The line is
$T_{c}=aH_{c2}(0)^{1/2}$ with $a=5.3KT^{-1/2}$.}
\label{fig3}
\end{figure}
 Given the unexpected empirical relation (\ref{eq5}), the evidence for d-wave
BCS behavior in the temperature dependence of $\rho _{s}\left( T\right) $,
and the resulting irrelevance of phase fluctuations, it is suggestive to
explore the BCS scenario further. Considering the observable $O\left(
T\right) $ with critical amplitude $O_{0}$ and $O\left( T=0\right) =O(0)$,
including the gap $\Delta $, the correlation length $\xi $ and the upper
critical field $H_{c2}$, the critical amplitudes should scale as \cite{book,fetter}

\begin{eqnarray}
T_{c} &\propto &\Delta _{0}\propto \xi _{0}^{-1}\propto H_{c20}^{-1/2}
\notag \\
&\propto &\Delta \left( 0\right) \propto \xi \left( 0\right) ^{-1}\propto
H_{c2}\left( 0\right) ^{-1/2},  \label{eq8}
\end{eqnarray}%
Here, $\lambda \left( T\right) =\lambda _{0}t^{-1/2}$, $\Delta \left(
T\right) =\Delta _{0}t^{1/2}$, $\xi \left( T\right) =\xi _{0}t^{-1/2}$, $%
H_{c2}=H_{c20}t\propto 1/\xi ^{2}$, where $t=1-T/T_{c}$. These scaling
relations follow from $\Delta \left( T\right) \propto 1/\xi \left( T\right)
\propto H_{c2}\left( T\right) ^{-1/2}$, and $\Delta \left( 0\right) \propto
\Delta _{0}\propto T_{c}$.

 In order to clarify their consistency with experimental facts, we need the $%
T_{c}$ dependence of $H_{c2}$ and the gap $\Delta $. In Fig.\ref{fig3}a we
depicted $T_{c}$ versus $H_{c2}(0)$ taken from Y. Wang and H.-H. Wen derived
from specific heat measurements on single crystals \cite{wangwen}. Although
the data are rather sparse the flow to the QSM transition is revealed, and
consistency with the BCS scaling law (\ref{eq8}) can be anticipated.
Consistency also emerges from Fig.\ref{fig3}b, depicting the
magnetoresistance measurements of Rourke \textit{et al} \cite{rourke} on
single crystals .  Quantitative agreement with BCS theory for a d-wave
superconductivity in overdoped cuprates stems from heat transport
measurements in Tl$_{2}$Ba$_{2}$CuO$_{6+\vartheta }$ \cite{proust}.

\begin{figure}[htb]
\includegraphics[scale=0.4]{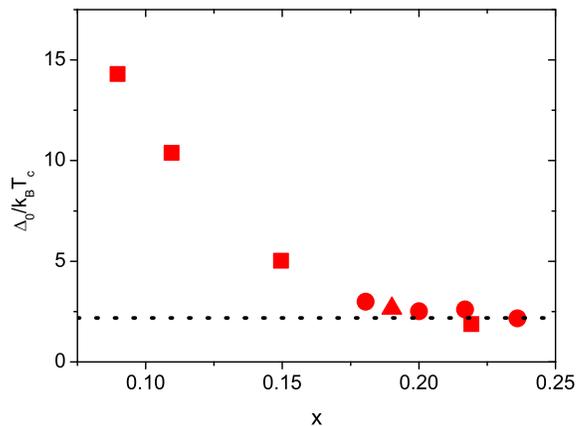}
\caption{Gap $\Delta \left( 0\right) /k_{B}T_{c}$ versus $x$ for La$_{2-x}$Sr$_{x}$CuO$_{4}$ 
taken from Wang \textit{et al} \cite{wang}. The dotted line
marks the limiting d-wave BCS value $\Delta \left( 0\right) /k_{B}T_{c}=2.14$. 
Note that for $x\gtrsim 0.19$ the data are consistent with the scaling
form (\ref{eq8}).}
\label{fig4}
\end{figure}

Fig. \ref{fig4} shows the doping dependence of the gap in terms of $\Delta
\left( 0\right) /k_{B}T_{c}$ versus $x$ for La$_{2-x}$Sr$_{x}$CuO$_{4}$ \
taken from Wang \textit{et al} \cite{wang}., derived from low temperature
specific heat measurements on single crystals. The dotted line marks the
d-wave BCS value $\Delta \left( 0\right) /k_{B}T_{c}\simeq 2.14$. Note that
for $x\gtrsim 0.19$ the data are consistent with the scaling form (\ref{eq8}%
) and in addition with d-wave BCS (weak coupling) superconductivity. More
compelling evidence for this scenario stems from heat transport measurements
in Tl$_{2}$Ba$_{2}$CuO$_{6+\vartheta }$ \cite{proust}.
Potential candidates appear to be the nonlocal effects on the penetration
depth of clean d-wave superconductors explored by Kosztin and Legget \cite%
{kosztin}. Taking the nonlocal effects into account they obtained for a
specular boundary the relation%
\begin{equation}
\frac{\lambda _{spec}\left( 0\right) }{\lambda _{L}\left( 0\right) }=1+\frac{%
\pi \sqrt{2}}{16}\frac{\xi \left( 0\right) }{\lambda _{L}\left( 0\right) },
\label{eq12}
\end{equation}%
where $\lambda _{L}\left( 0\right) $ is the zero temperature London
penetration depth. Considering the limit $\xi \left( 0\right) /\lambda
_{L}\left( 0\right) $ $>>1$ they expected no significant corrections at zero
temperature. This differs from the present case where the QSM transition is
approached, $\xi \left( 0\right) \propto 1/T_{c}$ tends to diverge and the
limit $\xi \left( 0\right) /\lambda _{L}\left( 0\right) $ $>>1$ is attained.
In this case the nonlocal corrections dominate and Eq. (\ref{eq12}) reduces,
in accordance with the empirical relation ((\ref{eq6}), to $\lambda
_{spec}\left( 0\right) \propto \xi \left( 0\right) \propto 1/T_{c}$. To
estimate the $T_{c}$ regime where this limit is reached we calculate the $%
T_{c}$ dependence of $\xi \left( 0\right) $. With $\xi \left( 0\right)
^{2}=\Phi _{0}/\left( 2\pi H_{c2}\left( 0\right) \right) $ and $%
T_{c}=aH_{c2}(0)^{1/2}$ with $a=5$ KT$^{-1/2}$ (see Fig.\ref{fig3}) we
obtain \ $\xi \left( 0\right) \simeq 900/T_{c}$ A with $T_{c}$ in K.
Consequently $\xi \left( 0\right) /\lambda _{L}\left( 0\right) $ $>>1$ is
fulfilled if $\lambda _{L}\left( 0\right) <<900/T_{c}$ A, where $\lambda
_{L}\left( 0\right) $ is fully determined by the properties of the Fermi
surface.

 In summary, we analyzed the temperature dependence of the superfluid
stiffness of selected overdoped La$_{2-x}$Sr$_{x}$CuO$_{4}$ thin films,
using the data of Bo\v{z}ovi\'{c} \textit{et al}. \cite{bosovic}. The
temperature dependence did not exhibit any sign of the expected KTB
transition. We observed remarkable consistency with a linear temperature
dependence, pointing to d-wave BCS behavior. On the other hand, we have
shown that the critical amplitude $\rho _{s0}$ and the zero temperature
counterpart $\rho _{s}(0)$ adopt essentially the same $T_{c}$ power law
dependence. This contradicts the standard result, whereupon $\rho _{s}(0)$
remains finite and is determined by Fermi surface properties. Exceptions
include dirty limit superconductors, where  $\rho _{s}(0)\propto T_{c}$.
Moreover we have shown that the $T_{c}$ of $\Delta \left( 0\right) $ and $%
H_{c2}\left( 0\right) $ are consistent with the expected BCS behavior, as it
should be, because both are proportional to some power of the correlation
length $\xi \left( 0\right) $. Noting that the correlation length is the
essential length scale the measurements of Bo\v{z}ovi\'{c} \textit{et al}.
\cite{bosovic} imply that $\rho _{s}(0)\propto \rho _{s0}\propto
\xi \left( 0\right) ^{-2}\propto \xi _{0}^{-2}$ holds by approaching the TSM
and QSM transitions down to 5 K. We observed that potential candidates appear to be
the nonlocal effects on the penetration depth of clean d-wave
superconductors explored by Kosztin and Legget \cite{kosztin}. It should be
kept in mind, that closer the QSM transition ($T_{c}=T=0$) quantum
fluctuations are expected to modify the outlined  scaling behavior and
uncover the difference between bulk and thin film samples.

Acknowledgements
Karl Alex M\"{u}ller I am grateful for our friendship dating back to
discussions on high temperature superconductivity in metallic hydrogen in
1970.

\end{document}